\documentclass[10pt,journal,compsoc]{IEEEtran}

\ifCLASSOPTIONcompsoc
  \usepackage[nocompress]{cite}
\else
  \usepackage{cite}
\fi

\ifCLASSOPTIONcompsoc
  \usepackage[caption=false,font=footnotesize,labelfont=sf,textfont=sf]{subfig}
\else
  \usepackage[caption=false,font=footnotesize]{subfig}
\fi
\ifCLASSOPTIONcompsoc
  \usepackage[nocompress]{cite}
\else
  \usepackage{cite}
\fi
\usepackage{url}
\usepackage[pdftex]{graphicx}
\usepackage{algorithmic}
\usepackage{colortbl}
\usepackage{etoolbox}
\usepackage{tikz}
\newcommand{\marker}[1]{\tikz{\node[draw=red,fill=red,circle,minimum
width=0.2cm,minimum height=0.2cm,inner sep=0pt,text=white, font=\bfseries] at (0,0) {#1};}}
\graphicspath{figure/}
\definecolor{lightgray}{gray}{0.9}
\definecolor{white}{gray}{1}
\begin{document}

\title{ASRPU: A Programmable Accelerator for Low-Power Automatic Speech Recognition\title{ASRPU: A Programmable Accelerator for Low-Power Automatic Speech Recognition}

}

% ~\IEEEmembership{Fellow,}
\author{Dennis~Pinto,
        Jose-María~Arnau,
        and~Antonio~González,~\IEEEmembership{Fellow,~IEEE}% <-this % stops a space
\IEEEcompsocitemizethanks{\IEEEcompsocthanksitem D. Pinto, JM. Arnau and A. González are with the Department of Computer Architecture, Universitat Politècnica de Catalunya, Barcelona, Spain.\protect\\
E-mail: \{dpinto, jarnau, antonio\}@ac.upc.edu}% <-this % stops an unwanted space
}

%{Shell \MakeLowercase{\textit{et al.}}: Bare Demo of IEEEtran.cls for Computer Society Journals}

\IEEEtitleabstractindextext{%
\begin{abstract}
The outstanding accuracy achieved by modern \textit{Automatic Speech Recognition (ASR)} systems is enabling them to quickly become a mainstream technology. ASR is essential for many applications, such as speech-based assistants, dictation systems and real-time language translation. However, highly accurate ASR systems are computationally expensive, requiring on the order of billions of arithmetic operations to decode each second of audio, which conflicts with a growing interest in deploying ASR on edge devices. On these devices, hardware acceleration is key for achieving acceptable performance. However, ASR is a rich and fast-changing field, and thus, any overly specialized hardware accelerator may quickly become obsolete. 

In this paper, we tackle those challenges by proposing ASRPU, a programmable accelerator for on-edge ASR. ASRPU contains a pool of general-purpose cores that execute small pieces of parallel code. Each of these programs computes one part of the overall decoder (e.g. a layer in a neural network). The accelerator automates some carefully chosen parts of the decoder to simplify the programming without sacrificing generality. We provide an analysis of a modern ASR system implemented on ASRPU and show that this architecture can achieve real-time decoding with a very low power budget.
\end{abstract}          
\begin{IEEEkeywords}
Parallel Architectures, Machine Learning, Automatic Speech Recognition, Parallel Architectures, Machine Learning, Automatic Speech Recognition, Real-time On-edge ASR.
\end{IEEEkeywords}}

% Abstract & Introduction: 3 pages
\maketitle
\IEEEdisplaynontitleabstractindextext
\IEEEpeerreviewmaketitle

\IEEEraisesectionheading{\section{Introduction}\label{sec:introduction}}

\begin{figure}[t]
    \centering
    \includegraphics[width=1\linewidth]{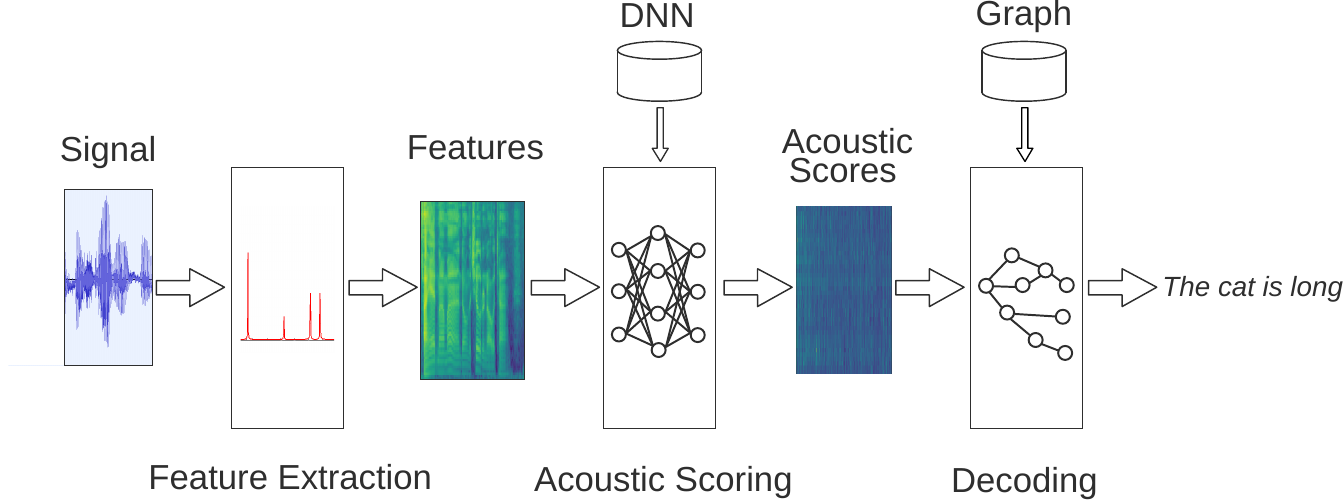}
    \caption{Overall diagram of an Automatic Speech Recognition system.}
    \label{fig:asr_diagram}
\end{figure}

\IEEEPARstart{A}{utomatic} speech recognition (ASR) consists of processing an audio signal (\textit{utterance}) to obtain a written transcription. Figure~\ref{fig:asr_diagram} shows the overall ASR process. First, the audio signal is broken down into overlapping fragments and transformed into a sequence of \textit{feature frames}. Then, each of these frames is classified into acoustic tokens by an \textit{Acoustic Model (AM)}. For most systems, the AM is a \textit{Deep Neural Network (DNN)} whose output is a vector of probabilities over phonetic units. The last stage, \textit{decoding}, generates a transcription from the acoustic scores. The simplest approach for obtaining a transcription consists of selecting the phonetic unit with the highest probability on each frame. However, using more sophisticated approaches, e.g. integrating a lexicon and a language model, generally result in better accuracy. When the ASR system includes a lexicon or a language model, the best-scoring transcription cannot be obtained by simply taking the single best-scoring phonetic unit on each frame. Instead, a search algorithm, such as \textit{Viterbi Beam Search}, traverses the graph of possible transcriptions (\textit{lattice}) to find the sequence of phonetic units with the best overall score.

ASR systems are key components in many game-changing technologies such as automatic language translation~\cite{kr2019towards} and virtual assistants~\cite{cortana, alexa, siri} as well as many others like dictation~\cite{poder2018speech}, automatic captioning~\cite{evain2020towards}, hands-free computing~\cite{sahadat2018comparing, vskraba2019development}, pronunciation evaluation for language learning~\cite{arora2018phonological} and many more.

\begin{figure}[t]
    \centering
    \includegraphics[width=1\linewidth]{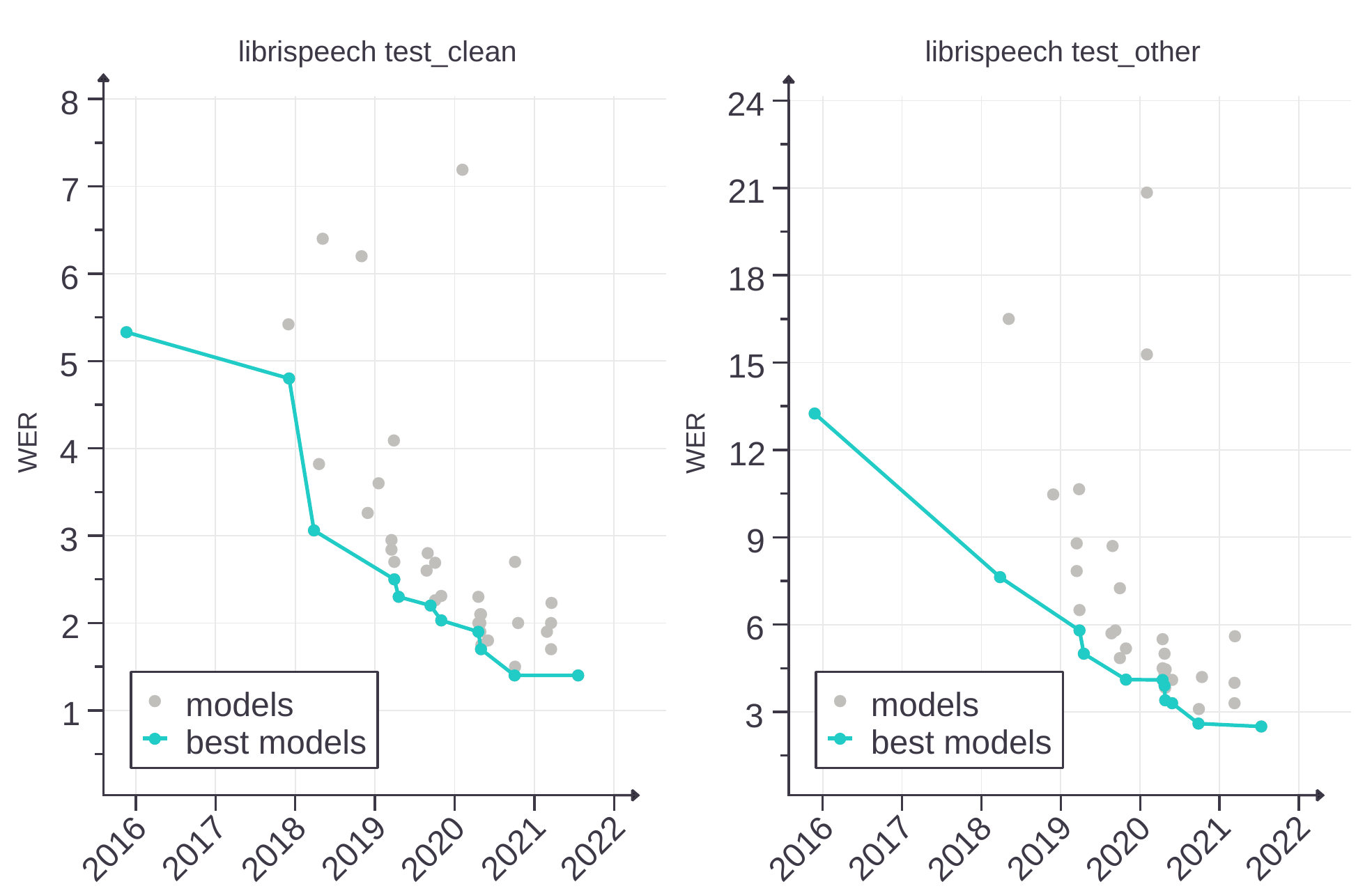}
    \caption{\textit{Word Error Rate (WER)} of different ASR systems on the librispeech test\_clean and test\_other benchmarks \cite{asr_wer}}
    \label{fig:summary_asr}
\end{figure}

To a higher or lesser degree, all of those applications require that the ASR system provides high-quality transcription while allowing the user to speak naturally, something that past ASR systems could not deliver. Modern ASR systems, however, are starting to excel at that. Figure \ref{fig:summary_asr} illustrates how fast ASR accuracy has improved over the last few years. The plot shows the \textit{Word Error Rate (WER)} of different ASR systems, as reported in various papers published between 2016 and 2021 for librispeech, a large-vocabulary, multi-user continuous speech recognition benchmark. The outstanding improvement is apparent by comparing a 2016 system, DeepSpeech2~\cite{amodei2016deep}, with the best 2021 system~\cite{zhang2020pushing}. In just 5 years, the WER was reduced from 5.33\% to 1.4\% in test\_clean, the basic librispeech task, and from 13.5\% to 1.7\% in test\_other, a librispeech task containing only challenging utterances. For reference, the WER of humans is estimated to be around 5\% for test\_clean and 13\% for test\_other~\cite{amodei2016deep}.

%%% ==> What is the problem?

This level of accuracy opens the door for many mainstream uses, as witnessed by the proliferation during the last decade of consumer products based on ASR. It comes, however, at the cost of performing inferences with huge models that require on the order of billions of arithmetic operations per second of speech and expensive searches in large graphs, such as lexicon and language models graphs. 

%For instance, in \cite{han2020contextnet}, authors propose \textit{ContextNet}, a DNN for ASR. It achieves 2.4\% (WER) on librispeech test\_clean, a speech recognition benchmark, when configured with 31.4M parameters and 2.1\% WER when scaled up to 112.7M parameters. Similarly, The \textit{Conformer} network \cite{gulati2020conformer} achieves 2.3\% and 2.1\% WER on the same benchmark with 30.7M and a 118.8M parameters, respectively. In \cite{synnaeve2019end}, authors use three types of networks: a 500M parameter \textit{ResNet} that achieves 2.67 WER, a 500M \textit{Time-Depth Separable (TDS)} network that attains 2.35 WER and a 296M parameter \textit{Transformer} network that achieves 2.25\% WER. Baevski~et.~al~\cite{baevski2020wav2vec} train a 317M parameter Transformer network that reaches 2.2\% WER. Chan~et.~al~\cite{chan2021speechstew} achieve 2\% WER with a 100M parameter network, Xu~et.~al~\cite{xu2021self} achieve 1.7\% WER with a 300M parameter network and Zhang~et.~al~\cite{zhang2020pushing} obtain 1.5\% WER with a 1B parameter network.

Another challenge for on-edge ASR comes from the wide diversity of ASR systems \cite{bhattacoustic}. ASR systems come in one of two major flavors: Hybrid DNN-HMM systems \cite{wang2020transformer, luscher2019rwth, xu2018neural} and End-to-End systems \cite{zhang2020pushing, xu2021self, park2020improved, baevski2020wav2vec, synnaeve2019end}. The former relies on a statistical model called \textit{Hidden Markov Model}~\cite{juang1991hidden} to identify and chain together individual spoken units, and a DNN to generate the inputs required by the HMM from the audio frames, whereas the latter consists of a Deep Neural Network (DNN) that classifies the audio frames into spoken units, generating a valid transcription directly. Additionally, End-to-end systems are often sub-classified depending on whether they are \textit{Connectionist Temporal Classification (CTC)} \cite{graves2006connectionist} or \textit{seq2seq}~\cite{chan2016listen} systems.

Furthermore, ASR systems are evolving rapidly. The impressive drop in transcription errors showcased in figure~\ref{fig:summary_asr} is possible thanks to the abundant innovations proposed during the last few years. ASR systems are constantly changing to incorporate these innovations. Consequently, any overly specialized chip to accelerate ASR will likely become obsolete at once.

The consequence of these three factors: high computational cost, vast heterogeneity and fast pace of innovation, is that ASR is usually performed on servers rather than on edge devices~\cite{cheng2020task}. Edge devices, such as smartphones and smart appliances are often ill-equipped to perform highly accurate ASR within reasonable latency. In contrast, servers provide more than enough computing power for the task. Furthermore, hardware and software in servers can be easily updated and so the companies can always guarantee state-of-the-art ASR to their clients. This is more challenging if ASR is deployed on edge devices. Edge devices generally require hardware acceleration to provide highly accurate decoding within reasonably latency \cite{Pinto2020Design, yazdani2016ultra, tabani2017ultra}, but too specialized hardware is likely to become obsolete rather quickly, leaving users stuck with sub-par ASR until they update the device.

Despite the advantages of servers for ASR, on-edge ASR is the preferred solution for the long term. Service availability and low latency requirements are difficult or plain impossible to guarantee when ASR is provided as a cloud service. Even more important are privacy issues that arise when sending sensitive audio data to company-owned servers.

%%% ==> How do we solve the problem?
Acknowledging the clear advantages of ASR on the edge, we propose \textit{ASRPU}, a processing unit for ASR, to tackle the challenges of executing ASR on edge devices. This accelerator is built around a pool of general-purpose cores, which gives the programmer flexibility to write alternative ASR implementations and perform software updates and optimizations as needed. The pool of cores is supported by an ASR controller and a unit that sorts and prunes hypotheses to automatize as much of the ASR process as possible without removing much flexibility. The accelerator also contains a specialized memory hierarchy adapted to the memory requirements of ASR. As a result, ASRPU provides enough flexibility to implement most of the current (and probably future) ASR systems with a simple and comprehensive API, enabling real-time state-of-the-art ASR on the edge.

Even though there is a plethora of accelerators for DNNs~\cite{chen2016diannao, chen2016eyeriss, chen2019eyeriss, putic2018dyhard} and ASR~\cite{tabani2017ultra, yazdani2016ultra, Pinto2020Design, price2016energy, liu2018eera}, to the best of our knowledge, we are the first to focus on flexibility and propose a low-power chip design capable of supporting a wide range of ASR systems.

The rest of the paper is organized as follows: Section~\ref{sec:background} provides background on Automatic Speech Recognition, introducing most of the important concepts and algorithms related to ASR, including examples and alternative algorithms. Section~\ref{sec:architecture} is a detailed description of ASRPU. Section~\ref{sec:programming_model} describes the detailed implementation of a modern ASR system on ASRPU as an example. Section~\ref{sec:results} presents our estimations of the performance and power consumption of a low-power configuration of ASRPU executing the ASR system introduced in section~\ref{sec:programming_model}. Finally, section~\ref{sec:conclusion} contains a summary of the conclusions and a discussion on both the limitations of this work and possible directions for future work.
\section{Automatic Speech Recognition} \label{sec:background}
As previously stated, the purpose of ASR is to obtain a written transcription from an utterance. Most current ASR systems do it by following the same overall algorithm consisting of three steps: (1) Feature extraction, (2) Acoustic Scoring and (3) Decoding. Feature extraction receives the raw signal and generates a sequence of \textit{feature frames}. These feature frames represent the signal in a convenient way, making the identification of phonetic units (e.g. phonemes) easier. The acoustic scoring step receives the feature frames and, employing an acoustic model, identifies the phonetic unit contained within each of the feature frames. The acoustic model generates, for each feature frame, a probability distribution over phonetic units. The decoding step combines the acoustic scores with scores from other sources (such as language models) to generate transcription hypotheses. When all the acoustic score frames are incorporated in the hypotheses by the decoder, the overall best hypothesis is regarded as the final transcription for the input utterance.

Despite the homogeneity implicit in the previous description, there are many alternative approaches to implement those processes. Feature extraction usually consists of a \textit{Mel-Frequency Cepstral Coefficients (MFCC)} extractor preceded by a signal processing step that enhances certain characteristics of the signal and reduces noise. However, MFCCs are not the only features used for ASR, and signal pre-processing can be performed in a variety of ways. The second process, acoustic scoring, usually consists on a DNN inference, but there is a wide variety of DNNs used among ASR systems (e.g. TDNN~\cite{peddinti2015time}, TDS~\cite{DBLP:journals/corr/abs-1904-02619}, DeepSpeech~\cite{amodei2016deep}, LAS~\cite{chan2016listen} and Transformer~\cite{synnaeve2019end}) and each comes with its own peculiarities. Decoding too can be performed in a variety of ways. Hybrid systems, such as \cite{zhang2020pushing, povey2016purely} do it by traversing a complex graph, which contains transition probabilities between a heterogeneous set of symbols, such as tri-phones, phonemes and words; end-to-end systems rely on simpler graphs. For example, \textit{wav2letter}~\cite{baevski2020wav2vec} systems employ a tree structure of phonemes, word-pieces or words. This structure serves to limit the symbols to which each symbol can transition. However, this graph does not contain scores. Additionally, a language model graph or DNN can be included in the systems. The language model consists of a graph that contains scores that represent transition probabilities between words.

The following sections provide a more detailed description of the common components of ASR systems and the alternative algorithms.

\subsection{Feature Extraction}

\begin{figure}[t]
    \centering
    \includegraphics[width=1\linewidth]{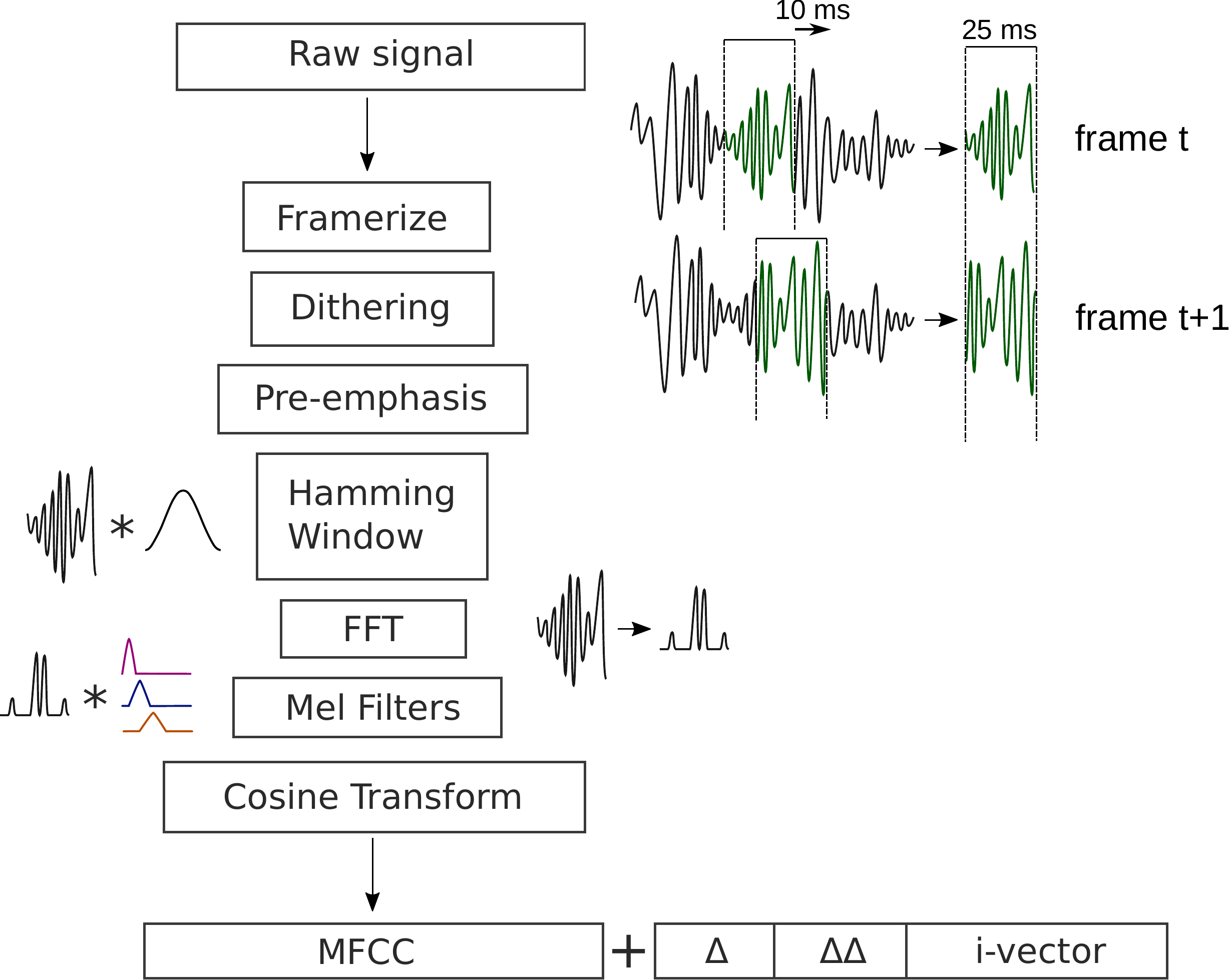}
    \caption{MFCC extraction algorithm.}
    \label{fig:mfcc}
\end{figure}

Feature Extraction is the first component of most ASR systems. It receives the raw signal (a sequence of amplitude values) and generates a sequence of feature vectors. These are usually MFCC features, which are computed roughly as follows (figure~\ref{fig:mfcc}): the signal is broken into overlapping frames, usually, 25ms frames shifted by 10ms, then, a Fourier Transform is applied on the signal segments to convert them into the frequency domain. The frames in the frequency domain are mapped to the \textit{mel scale} using overlapping triangular windows (\textit{mel filterbanks}). Usually, 80 of the resulting filterbanks compose a feature frame, whereas the rest are discarded. Finally, a cosine transform is applied to the element-wise log of the resulting frames. Dynamic features, such as delta and delta-delta can be appended to the feature vectors.

Additionally, some form of signal pre-processing can be applied to enhance certain frequencies or reduce noise. It is also common to rise the filterbanks to some power before applying the cosine transform in order to make the features more robust against noise. Other feature vectors can also be appended to the MFCC vector. For example, appending \textit{i-vectors}~\cite{dehak2010front} has been observed to improve decoding accuracy by providing speaker adaptation~\cite{rouvier2014speaker}.

Even though MFCC features are the most common, other types of features, such as PLP, are used in some systems and may provide additional benefits.

\subsection{Acoustic Scoring}
The feature frames produced by the feature extraction process are processed by the acoustic model to generate a sequence of score vectors, usually one vector per feature frame. Depending on the system, these vectors may contain probabilities or log-likelihoods. The set of acoustic tokens is defined by the ASR designer. Many recent systems use word pieces, but characters and phonemes are not rare.

\begin{figure}[!t]
  \centering
  \subfloat[ResNet Block]{\includegraphics[width=1\linewidth]{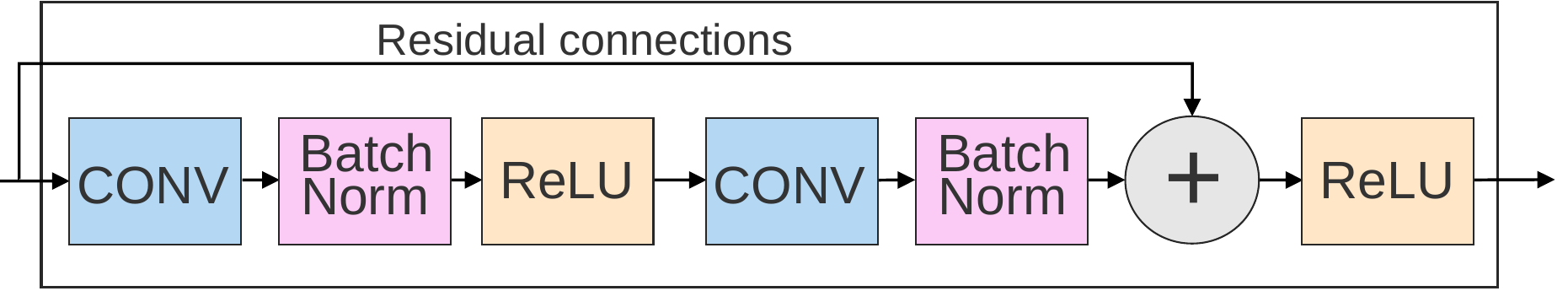}%
  \label{fig:resnet_diagram}}
  \hfil
  \subfloat[TDS Block]{\includegraphics[width=1\linewidth]{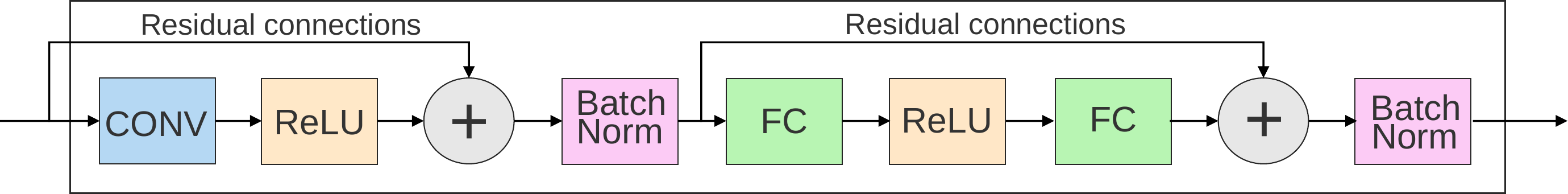}%
  \label{fig:tds_diagram}}
  \caption{Diagram of (a) a ResNet block and (b) a TDS block.}
  \label{fig:am_networks}
\end{figure}

Acoustic scoring is essentially a classification problem, so it is most commonly implemented as a DNN inference. However, Neural Networks is a broad and fast changing domain, with many architectures to chose from and new ones proposed every year. Common architectures include: \textit{Time Delay Neural Network (TDNN)}~\cite{peddinti2015time}, \textit{ResNet}~\cite{he2016deep} and \textit{Time-Depth Separable (TDS)}~\cite{hannun2019sequence} networks (figure~\ref{fig:am_networks}).

\subsubsection{Acoustic Tokens} \label{sec:background_acousticTokens}
The acoustic model classifies each feature vector by generating a vector of probabilities over Acoustic tokens. On Hybrid systems~\cite{povey2016purely, guglani2021dnn}, the acoustic tokens are HMM states. A WFST graph maps these tokens to phonemes and finally to words. End-to-End systems do not use HMM. Instead, on those systems, feature vectors are directly classified into phonemes, characters or word-pieces~\cite{zhang2020faster, zhang2020pushing}. The latter, word-pieces, are currently among the most popular type of acoustic tokens for end-to-end systems. They consist of arbitrary pieces of words obtained by an optimization algorithm that takes words from a text corpus and breaks them apart on different parts to minimize a target cost function. 

The selection of acoustic tokens influences the decoding phase. Word pieces and characters can be decoded with very simple decoding graphs or even without any graph. HMMs and phonemes, on the other hand, require a decoding graph to map these low-level tokens into characters or words.

\subsection{Decoding}
Once the sequence of acoustic scores is generated, they are consumed during the decoding stage to generate transcriptions hypotheses. This can be so simple as taking, for each frame, the phonetic unit with a higher score or as complicated as processing a large decoding graph that combines a Hidden Markov Model with additional graphs that model pronunciation and grammar.

\subsubsection{Decoding in Hybrid ASR}
Decoding in hybrid systems is performed by traversing a decoding graph while consuming acoustic score frames, looking for the best scoring path. The principal characteristic of hybrid systems is the inclusion of \textit{Hidden Markov Models (HMM)}~\cite{rabiner1989tutorial}, a set of weighted directed graphs representing acoustic units (usually tri-phones). HMMs are represented as \textit{Weighted Finite-State Transducers (WFST)}~\cite{mohri2004weighted} and then merged with a \textit{Lexicon}, i.e. a pronunciation model, and a \textit{Grammar}, i.e. a language model. Additionally, a \textit{Context dependency} graph is included in the mix to make the HMMs compatible with the Lexicon. These HMMs are trained on labelled utterances to learn how to align the HMM nodes to the acoustic vectors, i.e. which nodes should have a better score for which acoustic vectors. The acoustic model DNN is trained on aligned utterances to learn and generalize the alignments learnt by the HMM. The graph resulting from combining the HMM, context-dependency, lexicon and grammar graphs is called \textit{HCLG} graph and is the standard decoding graph for Hybrid ASR systems.

To traverse the HCLG graph, the \textit{Viterbi Beam Search} algorithm is used. This algorithm starts from a special start node. At each step, the algorithm checks all the nodes reachable from the set of active nodes, computing the score of the resulting paths (from the start node). The nodes with scores within the beam, a threshold computed from the score of the best node, will compose the active set for the next traverse step. The process finishes when all the acoustic vectors are consumed. To backtrack the best path once the algorithm ends, after each traverse step, for every active node, a pointer to the parent node is recorded. If a node was reachable from several parent nodes, all but the best scoring are discarded.

\subsubsection{Decoding in end-to-end ASR}
End-to-End ASR does not include an HMM to model acoustic units. Instead, the acoustic model DNN is trained from scratch to learn how to align the utterances. One of the consequences in this model is that tri-phones are not necessary, and consequently, neither are context-dependency nor lexicon. The acoustic model DNN in these systems processes feature vectors and generates scores over high-level phonetic units, such as characters or word-pieces.

The most straightforward algorithm consists in taking, for each frame, the best scoring phonetic unit. However, this approach often results in poor accuracy. Introducing a lexicon and a language model restricts the possible paths, leading to more compelling results.

Since the acoustic tokens are characters or word-pieces, there are no alternative paths to generate the same word (as opposed to a triphone representation, where converging paths are very common). Because of this, the lexicon can be efficiently represented with a tree structure of phonetic units. The path from the root to a leaf node contains a sequence of phonetic units that form a complete word.

\subsection{Streaming Decoding}
\textit{Streaming decoding} (sometimes called \textit{online decoding}) refers to the process of decoding the speech in real-time while is generated and, hence, not having the entire utterance available. In streaming decoding, frames are decoded one-by-one, or in a batch of enough input frames to generate one output. In contrast, in non-streaming, or offline decoding the entire input utterance is available at the start of the decoding process. This distinction is important because the difference in decoding latency among the two options may be significant. streaming decoding generates partial transcriptions in real-time with very low latency, whereas offline decoding will generate the complete transcription after the speaker has finished an utterance plus some delay.

If the ASR is executed locally (on-edge), streaming decoding is desirable in order to grant the best user experience, e.g. immediate feedback and early detection of transcription mistakes are likely important advantages from the point of view of the user. However, these advantages come at a cost. During streaming decoding, only a small number of input frames are available, which means that there is less potential for data reuse during DNN inference. This is critical during the execution of fully-connected layers. These layers usually contain a huge number of parameters, which are used just once per input frame. Consequently, the parameters from fully connected layers will have to be accessed from external memory many times, incurring higher energy costs.        % Background: 3 pages
\section{Architecture of ASRPU}\label{sec:architecture}

Despite the differences among ASR systems, most of them follow a similar overall algorithm. We leverage that to design an accelerator that provides enough flexibility to support the differences between them while automating wherever possible to speed up the ASR process and simplify the software implementation.

\begin{figure}[t]
    \centering
    \includegraphics[width=1\linewidth]{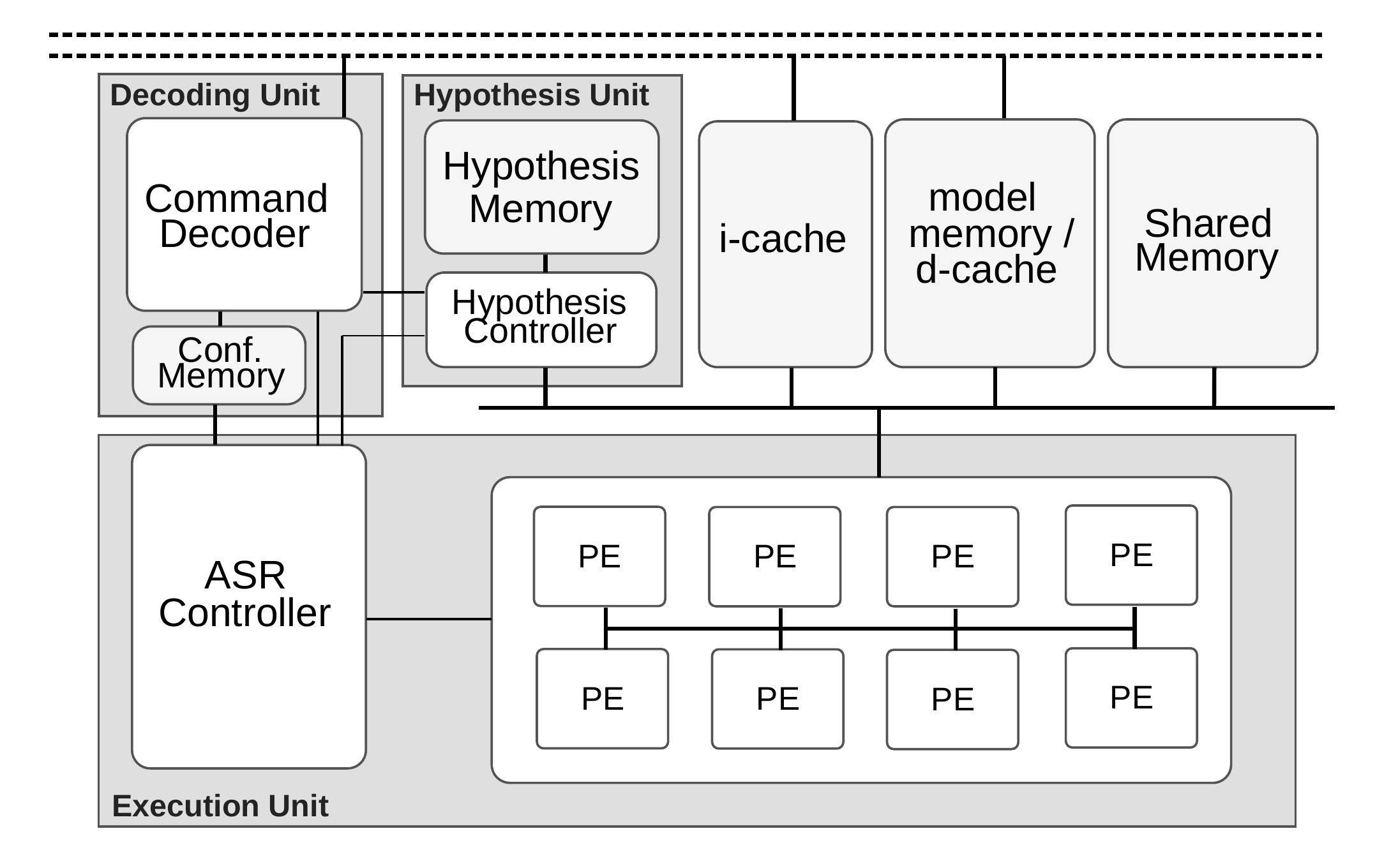}
    \caption{Architecture of ASRPU}
    \label{fig:arch_diagram}
\end{figure}

This section provides a detailed description of the architecture of ASRPU. The accelerator (Figure~\ref{fig:arch_diagram}) is divided into $3$ major blocks: \textit{Command decoder}, \textit{Execution unit} and \textit{Hypotheses unit}. The command decoder provides the interface to the accelerator via a set of commands. This includes commands to start an ASR decoding step, to finish decoding an utterance and to configure different parameters of the accelerator. The execution unit executes the program that implements the ASR system. This program is composed of small sub-programs, \textit{kernels}, written by the ASR designer to implement each part of the ASR system. The execution unit contains a pool of \textit{Processing Elements (PE)} to execute the code of the kernels. The hypothesis unit sorts and prunes transcription hypotheses. It also keeps them in memory from one decoding step to the following.

\begin{figure}[t]
    \centering
    \includegraphics[width=1\linewidth]{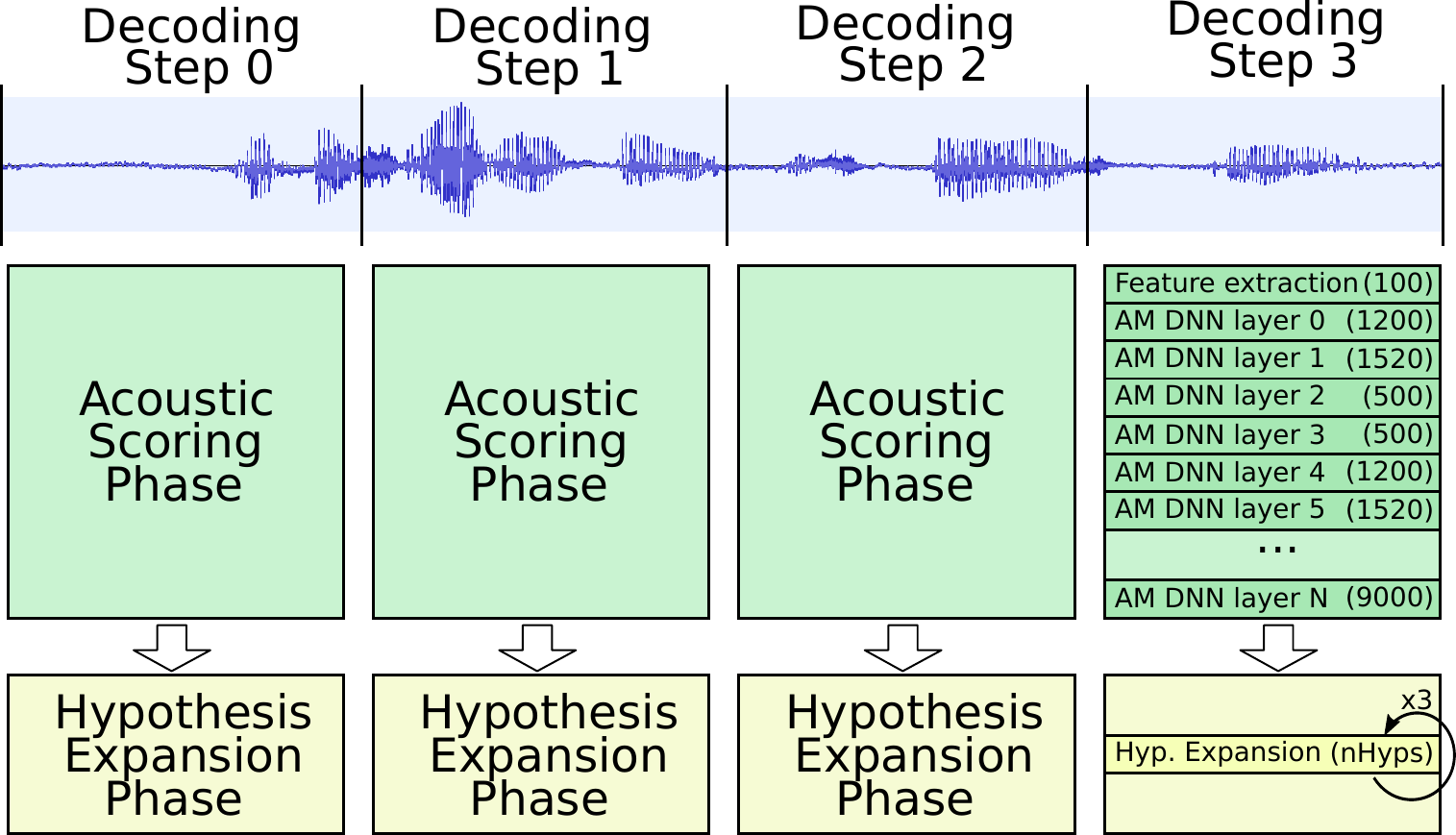}
    \caption{ASR process executed on ASRPU}
    \label{fig:overal_asr}
\end{figure}

%\begin{figure}[t]
%    \centering
    %\subfigure[Figure A]{\label{fig:a}\includegraphics[width=60mm]{example-image-a}}
    
%    \subfigure[]{\label{fig:overal_asr}\includegraphics[width=0.6\textwidth]{figure/asr_process_in_accel.pdf}}
%    \subfigure[]{\label{fig:asr_setup}\includegraphics[width=0.18\textwidth]{figure/asr_process_in_accel_setup_programs.pdf}}

%    \caption{(a) Overall decoding process in the accelerator. (b) The Acoustic Scoring Phase can optionally include \textit{setup} programs ($S_{fe}$, $S_0$, $S_1$,..., $S_N$) to configure the number of threads at run-time.}
%    \label{fig:three graphs}
%\end{figure}

\subsection{Decoding on ASRPU}
% Explain sequence of programs: setup, computation

Figure~\ref{fig:overal_asr} illustrates the overall process of decoding an utterance in ASRPU with an example ASR system. The decoding process in the accelerator is divided in \textit{Decoding Steps}. Each step decodes a portion of the signal, extracting feature frames, computing acoustic scores and finally expanding the hypotheses left from the preceding decoding step. We divide each decoding steps in two phases: (1) The \textit{Acoustic Scoring phase} and (2) the \textit{Hypothesis Expansion phase}.

As previously mentioned, a set of kernels implement every component of the ASR system. The acoustic scoring phase consist of the sequential execution of most of these kernels (except for the last one). These kernels implement the feature extraction algorithm and the acoustic model. The example of the figure shows a sequence of N+1 kernels executed within the acoustic scoring phase. These kernels are executed sequentially on the accelerator. However, they consist of parallel code. The execution of each kernel is carried on by the execution unit, which launches as many threads of the kernel code as required on the PEs. The number inside the parenthesis shows the number of threads required by each kernel. The first kernel implements feature extraction (which may include code for signal pre-processing) and requires 100 threads. Subsequent kernels implement each a layer of a DNN AM, each requiring a different number of threads. The last kernel requires $9000$ threads, which is entirely dependant on the implementation. In this example, the last kernel implements a DNN layer with $9000$ neurons. Each neuron computing the score for one of the $9000$ phonetic units modelled by the acoustic model. because of how the kernel is written, each thread computes a single neuron.

After the acoustic scoring phase concludes, ASRPU switches to the hypothesis expansion phase. During the hypothesis expansion phase, the accelerator executes only one kernel, \textit{Hypothesis expansion}. Each thread of this kernel is responsible of expanding a single hypothesis. The expansion of an hypothesis generally results in many output hypotheses, which are generated according to the specific decoding algorithm. Depending on the implementation, the acoustic scoring phase can generate one or more acoustic vectors. During hypothesis expansion, the accelerator executes the hypothesis expansion kernel once per acoustic vector. In the example system of the figure, the accelerator launches \textit{nHyps} threads (determined in run-time) of the hypothesis expansion kernel. The self-referencing arrow indicates that the kernel is executed three times. This number will also depend on the implementation. For example, the feature extraction kernel may extract three frames on each decoding step, resulting in three repetitions of the hypothesis expansion kernel. Some DNNs, particularly convolutional DNNs, apply \textit{sub-sampling} during acoustic scoring, meaning that they generate less acoustic vectors than feature frames. In this case, six feature frames will result in three acoustic vectors if the DNN AM apply a sub-sampling of two frames.

If ASRPU is integrated in an SoC that also contains a CPU, there may be an external process responsible of capturing the signal as it is produced. This process communicates with the accelerator, starting decoding steps after capturing enough values from the microphone.

\subsection{Setup Thread} \label{sec:hardware_dynamic_batch}
The ASR designer can include a special \textit{setup program} along each of the acoustic scoring and hypothesis expansion kernels. That is, each kernel is complemented with a setup program. This setup program is executed to completion in a single thread before the associated kernel can start executing. 

These setup programs provide the accelerator with greater flexibility. For example, the setup program for a specific kernel that implements a convolutional layer of a DNN can determine how many outputs can be computed from the available inputs and notify the hardware to launch the appropriate number of kernel threads so as to maximize data reuse. The setup program associated to the hypothesis expansion kernel can access the number of outputs generated by the acoustic scoring phase and notify it to the hardware so it executes the hypothesis expansion kernels as many times as necessary. In both cases, if the available inputs are not enough to compute even a single output, the setup thread can notify to the accelerator to stop the decoding step. The following section provides mores details about this process.

These setup program can also be used to manage the input and output buffers of the kernels in shared memory. Each kernel will generally read inputs from an input buffer and store outputs in an output buffer (in shared memory). Before executing each kernel, the associated setup thread will first determine the number of outputs that can be generated from the inputs available in the input buffer. Then, it will remove from the input buffer those inputs that can not be further reused and reserve space in the output buffer for the new outputs. Finally, before finishing, it notifies the hardware the required number of kernel threads. After the setup thread finishes, the accelerator launches the required number of kernel threads.

Another advantage of the setup threads is that they allow to reuse code among different kernels. Generally, DNNs contain many convolutional and fully-connected layers. The ASR designer can write a single convolutional and a single fully-connected parametric routines. All the convolutional and fully-connected layers can be configured to execute the same kernels and the associated setup threads will set the appropriate parameters in shared memory before executing the kernel. 

\begin{figure}[t]
    \centering
    \includegraphics[width=1\linewidth]{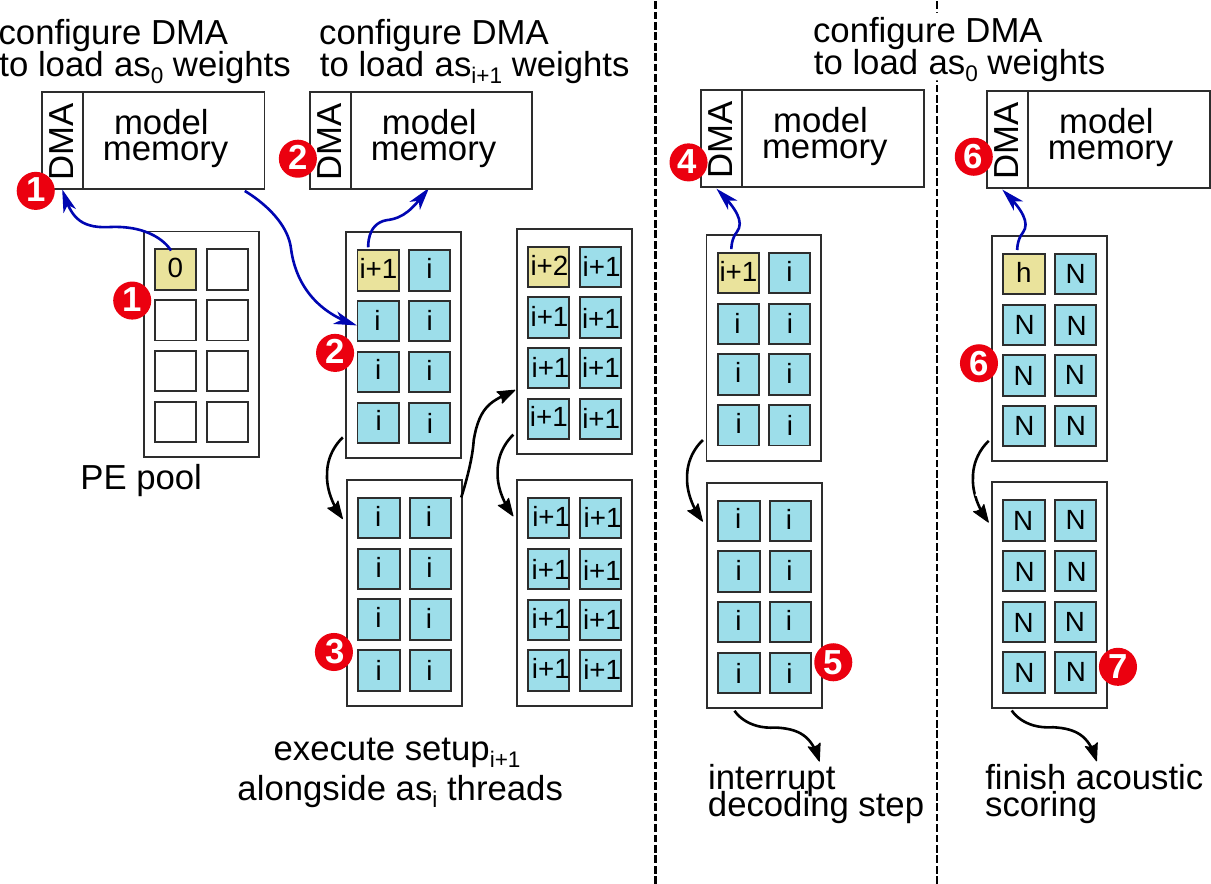}
    \caption{Threads in the PE pool}
    \label{fig:threads_in_pe_pool}
\end{figure}

\subsection{Execution Unit}
The execution unit consists of a pool of PEs and an \textit{ASR controller}. The ASR controller handles the overall decoding procedure. It first waits until the command decoder receives a new commit signal. At that moment, it starts a decoding step. First, the controller reads from the Configuration memory the address of the first setup program and configures a PE to execute it by setting its program counter. Once the setup thread finishes executing, it notifies to the ASR controller the required number of kernel threads. The ASR controller then starts dispatching kernel threads to idle PEs. Every time a PE becomes idle, it notifies the ASR controller, which reacts by dispatching a new thread to the PE, until there are no more threads to dispatch. When the last thread finishes, the ASR controller repeats the same procedure for the subsequent kernel.

As mentioned in section~\ref{sec:hardware_dynamic_batch}, if a setup thread returns a value of zero, the ASR controller stops the decoding step. This is meant to be used when a program is not ready to be launched, usually when there are not enough inputs to compute even a single output. For example, a convolutional layer with a window of ten frames will check during setup time (during the execution of the setup thread) how many inputs there are available, computing and returning an appropriate number of threads. If there are less than ten inputs, it will return zero notifying the ASR controller to stop the decoding step.

After all the programs in the Acoustic Scoring sequence have been executed, the decoder starts the hypothesis expansion phase. It first accesses the number of active hypotheses, provided by the hypothesis unit, and launches a thread for each active hypothesis. These threads will execute the code from the hypothesis expansion kernel. The setup thread of the hypothesis expansion kernel will determine how many outputs were generated by the acoustic scoring phase and notify the ASR controller to execute the hypothesis expansion kernel that number of times.

Figure \ref{fig:threads_in_pe_pool} shows how the different threads are scheduled in the PE pool during acoustic scoring. Each square represents a PE executing a setup thread (yellow) or a kernel thread (blue). \marker{1} First, the setup thread of kernel $0$ is dispatched. It configures the DMA to load the model data for kernel 0 in model memory and waits for it to finish. \marker{2} The execution of the following kernels ($as_i$ in the figure) starts by dispatching the setup thread for the next kernel ($as_{i+1}$) alongside the kernel threads of $as_i$. \marker{3} The ASR controller keeps dispatching $as_i$ threads until the kernel is completely executed. If a setup thread determines that the corresponding thread cannot be launched \marker{4}, it will notify the controller. Additionally, it can pre-fetch the model data for kernel $0$ to skip step \marker{1} during the next decoding step. After the current kernel finishes \marker{5}, the controller will interrupt the decoding step and wait for the next decoding command, which will start a new decoding step from \marker{1} or \marker{2}, depending on whether the model data for kernel 0 is pre-loaded or not. \marker{6} The setup for the hypothesis expansion phase is launched alongside the threads for the last acoustic scoring kernel. Finally, when all the threads for the last acoustic scoring kernel finish \marker{7}, the accelerator ends the acoustic scoring phase.

\subsection{Processing Elements}
\begin{figure}[t]
    \centering
    \includegraphics[width=0.5\linewidth]{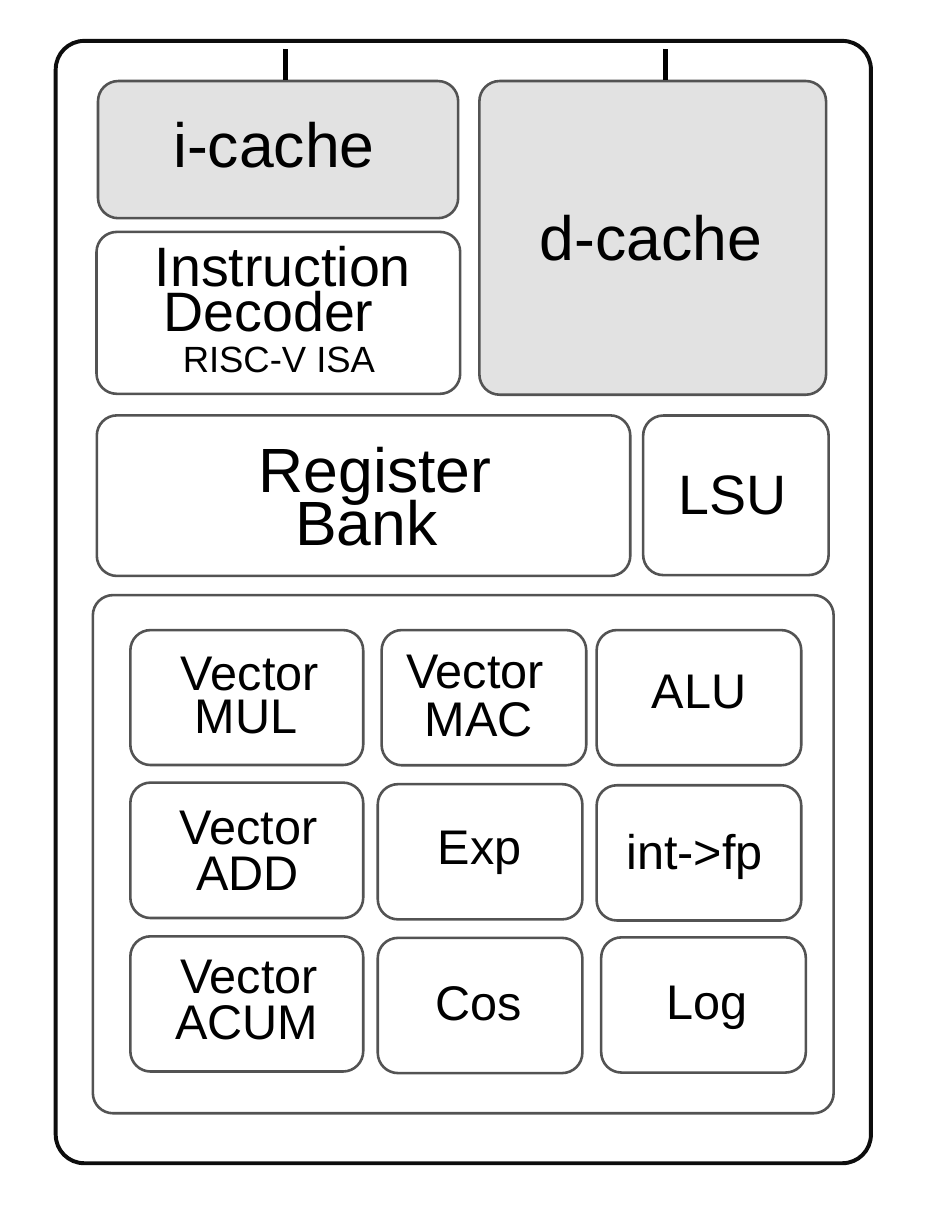}
    \caption{Processing Element (PE)}
    \label{fig:pe_diagram}
\end{figure}

The \textit{Processing Element (PE)} pool contains a number of programmable and independent PEs (i.e. cores). Each PE, shown in figure \ref{fig:pe_diagram}, implements a general-purpose RISC-V ISA. 

The ISA includes extensions for additional operations, such as a vector \textit{Multiply and Accumulate (MAC)}. This operation receives three operands, the first operand is a 32-bit value that carries the accumulation between MAC operations. The other two operands are vectors of 8-bit values. These operands are multiplied element-wise and accumulated. The result is added to the first operand. It also includes vector multiplication and additions, along with especial function units to compute logarithms, exponential and cosine functions, usually required during feature extraction, the activation function in neural network layers and for the computation of the hypothesis score during hypothesis expansion. Each PE contains data and instruction caches. These are regular caches managed by the hardware. Each PE also contains a register bank with 2 sets of registers: 32-bit floating-point registers, which are used as operands for the FP ALU and the special function units, and vectors of 8-bit values used as operands for the vector operations. PEs are connected to the Hypothesis unit, the shared memory and the shared caches through a bus. Another bus connects all the PEs to the ASR controller. This bus is used by the ASR controller to configure the PEs and by the PEs to notify values to the ASR controller.

\subsection{Hypothesis unit}
The hypothesis unit contains a hypothesis memory and a controller. During any decoding step, the active hypothesis and the newly generated hypothesis reside inside the hypothesis memory. This unit is connected to the internal bus and accessed via a special memory address from the PEs. Hypothesis Expansion threads send hypotheses to the hypothesis controller. Each hypothesis is a data structure with some fields. These fields include a hash to identify the hypothesis, the hypothesis score, and others defined by the programmer. These can include a backlink, pointers to data structures (e.g. to a node in the decoding graph) or a token id, for example.

Hypothesis expansion threads access hypotheses from this unit and send back the newly generated hypotheses. The hypothesis unit sorts and prunes them according to their score field and the beam score. The score beam is configured beforehand via configuration commands.

\subsection{Memory Hierarchy}
Each PE contains data and an instruction cache. Outside the PE pool, there are shared instruction and data caches too. The global data performs two different functions. During acoustic scoring, this memory stores model weights that were pre-fetched beforehand. This maximizes data reuse and hides the latency to access external memory. During hypothesis expansion this scheme would not be of much use. The graph structures used by the hypothesis expansion algorithms are generally in the order of hundreds of MB or even GB, much larger than what is reasonable to store in a low power accelerator. Additionally, the threads access the graph structures following a random pattern. Consecuently, during the hypothesis expansion phase, the data cache acts as a regular LRU cache to leverage locality in the access to the graph structures.

ASRPU also includes a scratchpad memory (the \textit{Shared Memory}) that is be accessed from the threads executing in the PEs. This is were the kernel buffers and the kernel configuration parameters are stored, along with any other variables defined by the programmer.

\subsection{Command Decoder}
The command decoder is the interface between ASRPU and the rest of the units in the SoC. It provides a set of commands (table~\ref{tbl:api}). These commands include some to configure the kernels and setup programs for the ASR phases: \textit{ConfigureASR\_AcousticScoring, ConfigureASR\_HypExpansion} and commands to configure other parameters (\textit{ConfigureBeamWidth}). These configuration commands must be used to configure the decoder before any decoding begins. In addition to those, the API contains commands for run-time operations. \textit{DecodingStep} is to indicate the accelerator to decode a given signal. This signal is not decoded in isolation. Instead, it is appended to previously decoded signals, extending the current transcription hypotheses. Once the utterance is finished, \textit{CleanDecoding} can be called. This command notifies the accelerator that the utterance is finished. In response, the accelerator prepares itself to decode a new utterance, cleaning the hypotheses memory and resetting the internal state.

\begin{table*}[t]
  \caption{Commands provided by the command decoder}
  \label{tbl:api}
  \centering
  \begin{tabular}{lp{2cm}p{8cm}}
    \hline
    Command & Parameters & Description\\
    \hline
    ConfigureASR\_AcousticScoring & n\_kernel setup\_addr kernel\_addr & Configure kernel \textit{n} from the Acoustic Scoring phase. setup\_addr and kernel\_addr refer to the address in external memory pointing to the setup program and the kernel program, respectively. Should be called several times with incremental values of \textit{n} to configure all the kernels that implement the acoustic scoring phase.\\
\hline
    ConfigureASR\_HypExpansion & kernel\_addr & Configure the Hypothesis Expansion phase. kernel\_addr is the address in external memory pointing to the hypothesis expansion kernel.\\
\hline
    ConfigureBeamWidth & beam & Configure the \textit{beam width} used by the hypothesis unit to prune hypotheses during hypothesis expansion. \\
\hline
    CleanDecoding &  & Perform the neccesarry operations to start decoding a new utterance, such as removing the hypotheses from the hypothesis memory.\\
\hline
    DecodingStep & signal\_addr & Command the accelerator to start a decoding step. The accelerator will access the data located in signal\_addr in the external memory and perform a decoding step.\\
  \hline
\end{tabular}
\end{table*}      % Technique: 4 pages + 3 pages
\section{Case Study}\label{sec:programming_model}
To illustrate the versatility and simplicity of our programming model for ASR, we present the implementation of one of the end-to-end systems from \textit{wav2letter}. Features are 80-dim MFCCs computed from the pre-processed audio signal. The acoustic model is a TDS network, built from TDS blocks (figure~\ref{fig:tds_diagram}). It is mostly composed of fully-connected and convolutional layers. The activation function for most layers is a ReLU, followed by a layer normalization. Hypotheses are extracted by traversing a lexicon tree that includes all the words in the vocabulary and a mechanism to handle out-of-vocabulary words. Additionally, an n-gram language model provides language model scores for the hypothesis.

In our implementation, the kernels that implement the acoustic scoring phase will first pre-process the signal and generate the MFCC frames. Then, they perform inference with the TDS network to obtain the acoustic scores from each of the computed frames. On each hypothesis expansion execution, all the hypotheses are expanded one node forward in the lexicon tree, covering each reachable node. Every reached node in the tree is a new hypothesis for the following hypothesis expansion execution. Every time a hypothesis reaches a node in the lexicon tree that represents a word, a link in the n-gram language model graph is traversed. The n-gram graph contains language model scores that are included, along with the acoustic scores, the word penalty and others, in the computation of the hypothesis score. Hypotheses are compared based on this score and those with a lower score are pruned away by the hypothesis unit. In addition to the reachable nodes in the lexicon tree, hypothesis expansion generates two more hypotheses as part of the CTC algorithm: the blank symbol and the repetition.

\subsection{The Main Process}

The main process residing in the CPU orchestrates the overall decoding of utterances. It does so by calling commands from the API of the accelerator. Before the decoding starts, the main process configures the accelerator, setting all the necessary parameters, including the addresses in external memory of the kernels that implement the ASR system.

During decoding, the main process collects reading from the microphone. This example ASR system performs streaming decoding, meaning that every few milliseconds, the main process calls the submittSignal command to perform a decoding step on a partial signal. If this was not the case, the main process would capture the signal until the end of the utterance is reached and then call a submittSignal on the entire signal. 

%\begin{lstlisting}[language=C++, caption=Main process]
%AsrAccel.ConfigureAsr_AS(0, "FE_setup.bin", "FE_kernel.bin")
%for(int i = 1; i < numLayersDNN+1; i++) {
%  AsrAccel.ConfigureAsr_AS(i, as_setup_list[i], %as_kernel_list[i])
%}
%AsrAccelr.ConfigureAsr_HE("HE_setup.bin", "HE_kernel.bin")
%AsrAccel.CleanDecoding()

%while(true) {
%  bool isSpeech = False;
%  while (!isSpeech) {
%    microphone.waitUntilNumValues(signalBatchSize);
%    signal = microphone.GetSignal();
%    isSpeech = VAD.isSpeech(signalFragment);
%  }

%  while(isSpeech) {
%    AsrAccel.CommitSignal(signalFragment);
%    microphone.waitUntilNumValues(signalBatchSize);
%    signal = microphone.GetSignal();
%    isSpeech = VAD.isSpeech(signalFragment);
%  }

%transcription = AsrAccel.GetTranscription()
%AsrAccel.CleanDecoding()
%}

%\end{lstlisting}

\subsection{Acoustic Scoring}
The acoustic scoring phase executes the code that implements the feature extraction and the acoustic model. The acoustic scoring phase consists of a set of programs executed in sequence. In this case study, the first kernel performs signal pre-processing and extracts MFCC features frames from the input signal, whereas the rest implement each a layer of the TDS DNN. 

Before executing the feature extraction kernel, its setup thread is launched to check the size of the input signal and determine how many output frames can be computed from the available input. Then, it reserves memory for the output, marks the inputs as consumed and notifies the controller about how many main threads must be launched. The kernel threads then process the inputs to generate feature frames. Each thread computes a single feature frame, which means that for each output frame to compute, a feature extraction thread will be launched.

The subsequent kernels in the acoustic scoring phase implement the TDS DNN. It is implemented in a sequence of 79 kernels: 18 CONV, 29 FC and 32 LayerNorms, each preceded by its corresponding setup thread. To avoid repeating very similar code, the programs for CONV, FC and LayerNorm are parameterized. The setup thread sets the parameters in shared memory to the values corresponding to the current layer, which are accessed by the layer threads.

Each setup thread checks the number of inputs available (those generated by the previous layer), reserves memory for the outputs and notifies the ASR controller to launch the required number of threads for the layer program. Each CONV and FC thread compute a single neuron of the layer.

\subsection{Hypothesis Expansion}

The hypothesis expansion kernel implements the CTC decoding algorithm with lexicon and language model. Each thread processes a single hypothesis. The algorithm first accesses the node in the lexicon graph associated with the hypothesis, then, it traverses all the output links to access reachable nodes, generating a new hypothesis for each of them. Each hypothesis also contain a link to the language model graph, pointing to the last n-gram in the hypothesis. If a newly reached node in the lexicon graph represents a word, the hypothesis expansion thread will acess the node in the language model graph associated with the hypothesis and expand it one node further following the link that represents the newly added word. The node contains a language model score that is added to the score of the hypothesis. In addition to the hypotheses generated by traversing the lexicon graph, the CTC algorithm implemented in the hypothesis expansion threads require the generation of two more hypotheses: the first one obtained by appending to the hypothesis the last phonetic unit in the hypothesis to account for repetitions, which produce valid CTC paths. The other hypothesis is obtained by appending the \textit{blank unit}, which represents a frame that does not contain a phonetic unit.

%\begin{figure}[t]
%    \centering
%    \includegraphics[width=0.7\linewidth]{figure_temp/placeholder.pdf}
%    \caption{Example pseudocode for the hypothesis expansion.}
%    \Description{}
%    \label{fig:programming_model}
%\end{figure}
\section{Evaluation}\label{sec:results}
This section provides estimations on the performance of ASRPU when running the ASR system described throughout the previous sections. The goal of this section is to provide proof of the capacity of the proposed design to enable real-time ASR on very low-power devices. To that purpose, we studied a possible implementation for the TDS-based system described in the previous section and estimated its performance on the accelerator, configured to enable real-time ASR with that system. Furthermore, we estimate the power consumption and area footprint of that specific configuration. 

\subsection{Methodology and Scope}
To estimate performance, we count the number of instructions for each kernel. For example, a loop will usually consist of two instructions for the comparison and conditional jump, one instruction for the variable update and the instructions for the loop body, all multiplied by the average number of iterations. Additionally, one instruction is added for the variable initialization. We assume that every PE executes one instruction per cycle, so we divide the number of instructions by the clock frequency of the PEs to obtain execution time. 

To estimate chip area, we rely on several tools. Cacti for the memories, McPat for the PEs and the PE bus and Design compiler (using the \textit{Saed32hvt cell library}, which provides cell models at 32nm technology node) for the special function units. 

Peak power is estimated by adding together the leakage power and peak dynamic power for the logic units as obtained from the Power Compiler. The case of memories is slightly different. Cacti reports leakage power and access energy. In this case, we assume as peak power the scenario where all the ports are accessed once per cycle. Adding the energy consumed for those accesses, divided by the clock period gives the dynamic power, which we add, along with the leakage power given by Cacti, to the power consumed by the logic. 

This estimation, albeit not exhaustive, should provide a good approximation of the potential of the accelerator proposed in this work.

\subsection{Accelerator Configuration}

\begin{table}[t]
    \caption{Configuration parameters of the accelerator}
    \label{tbl:parameters}
    \centering
    \begin{tabular}{rl}
        \multicolumn{2}{c}{\textbf{ASR Unit}}\\
        \rowcolor{lightgray} Frequency & 500 MHz\\
        \rowcolor{white} Hypothesis Memory & 24 KB \\
        \rowcolor{lightgray}     I-Cache & 64 KB \\
        \rowcolor{white} Shared Memory & 512 KB \\
        \rowcolor{lightgray}     Model Memory / D-Cache & 1 MB \\
        \rowcolor{white} Num. PEs & 8 \\
        \\
        \multicolumn{2}{c}{\textbf{PE}}\\
        \rowcolor{lightgray}     PE i-Cache & 4 KB \\
        \rowcolor{white} PE d-Cache & 24 KB \\
        \rowcolor{lightgray}     MAC. vector size & 8 \\
    \end{tabular}
    
\end{table}

\begin{figure}[t]
    \centering
    \includegraphics[width=1\linewidth]{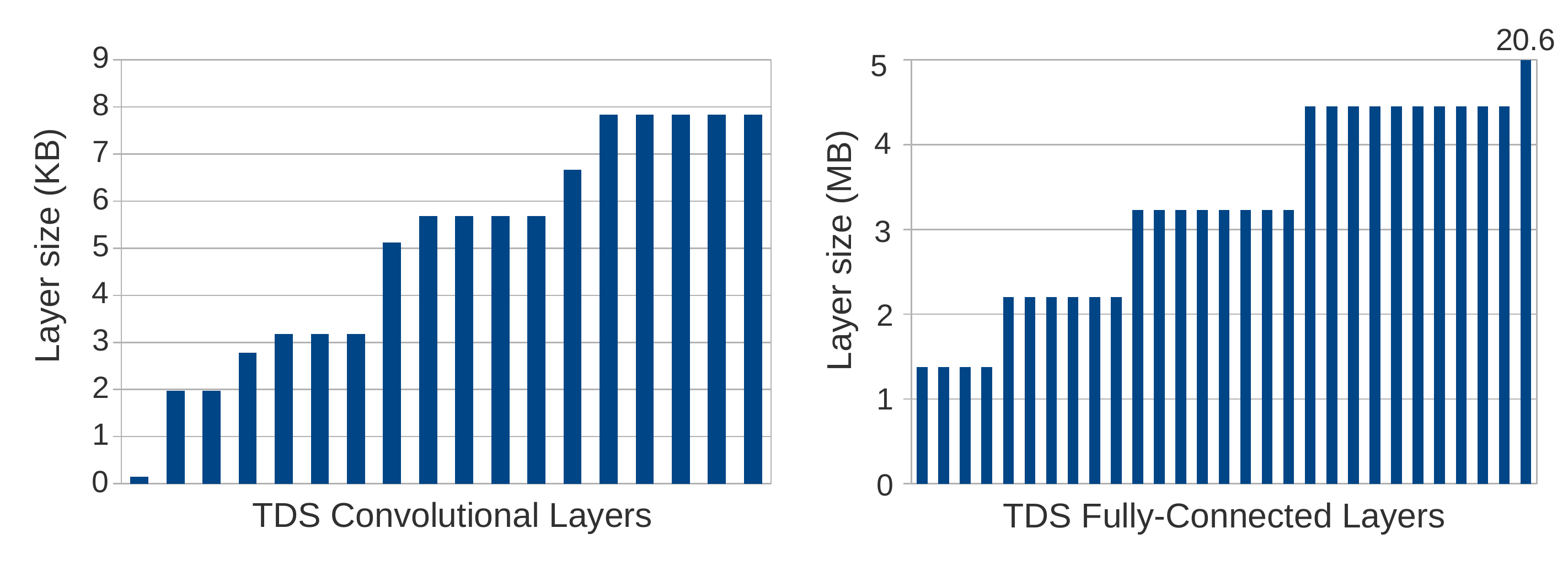}
    \caption{Size (KB) of each layer of the TDS DNN included in the ASR system. The left plot shows the convolutional layers whereas the right plot shows the fully-connected layers}
    \label{fig:tds_model_size}
\end{figure}

%%% TODO: Justify the parameters
Table~\ref{tbl:parameters} contains the details of the accelerator. This configuration was chosen to allow real-time ASR with the ASR system described in previous sections. Particularly, the number of PEs and the size of the memories was chosen to match the performance requirements. We include 8 PEs, each loaded with an 8-dim MAC unit, which allows us to exploit plenty of parallelism. The implemented algorithm stores about $275 KB$ of intermediate data in between decoding steps. It stores inputs for the convolutional layers. Due to the shifting input window used in convolutions, inputs are reused in several consecutive executions. We include $512 KB$ of shared memory to store these inputs and other temporal outputs that may be necessary to store if the decoding step is interrupted due to insufficient inputs for one of the kernels. 

We include $1 MB$ of model memory. During acoustic scoring, this memory is used for caching the DNN parameters and other model data. The size of the TDS network layers vary significantly (figure~\ref{fig:tds_model_size}). Convolutional layers fit in a few KB whereas most fully connected layers range in the MB. We solve this by trivially partitioning FC layers into several kernels, each less than $1 MB$. Given that each thread in our implementation of the FC kernels computes a neuron, we partition the layers in various kernels, each computing some of the neurons. For example, each of the first FC layers consists of $1200$ neurons with $1200$ inputs each, which results in $1.4 MB$ of model data. We divide each of these layers into 2 kernels, each computing $600$ neurons ($700 MB$).

\subsection{Area and Power}

\begin{figure}[!t]
  \centering
  \subfloat[Area and peak power by component]{\includegraphics[width=0.36\linewidth]{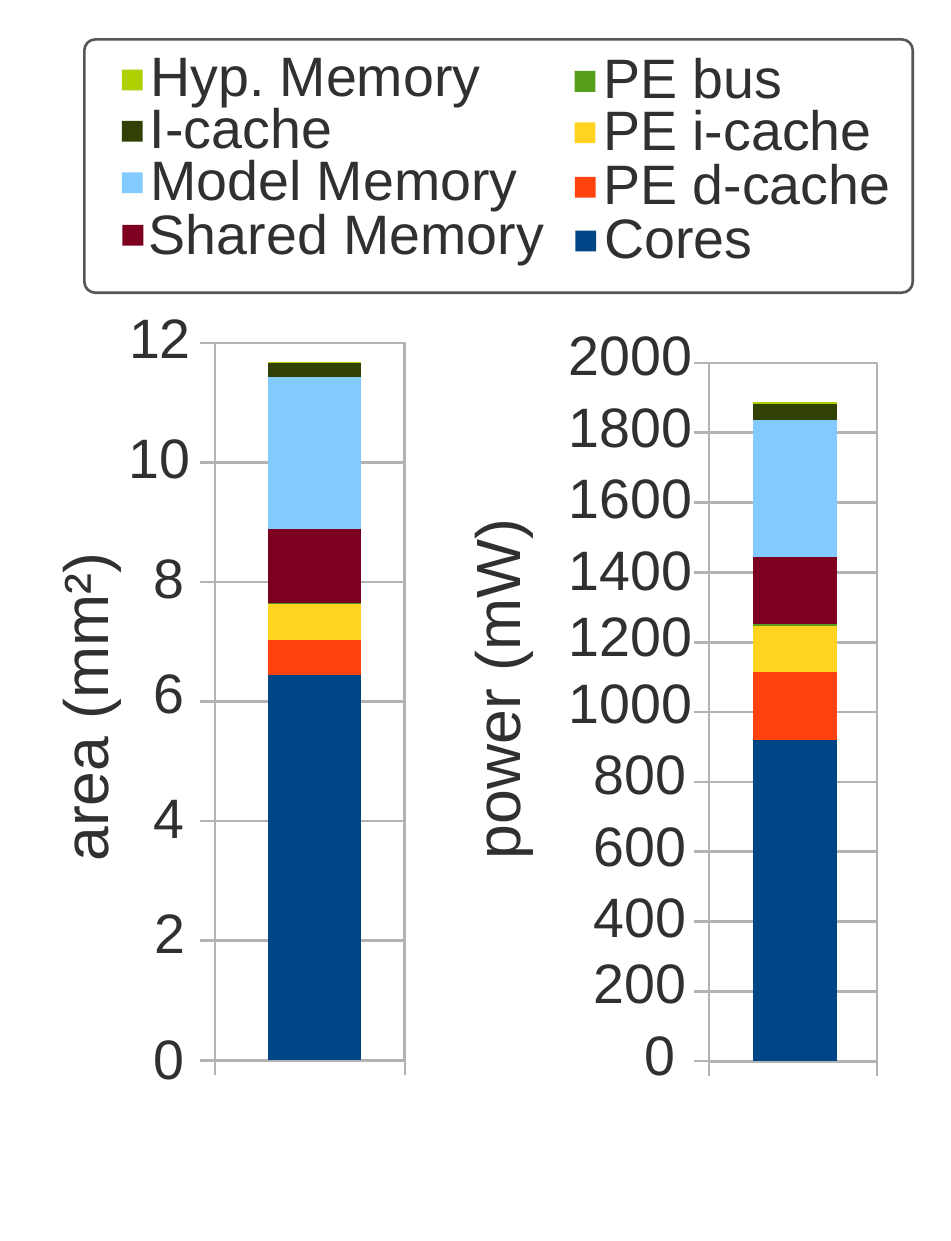}%
  \label{fig:accel_area_ppower_breakdown}}
  \subfloat[Dynamic-static power breakdown]{\includegraphics[width=0.64\linewidth]{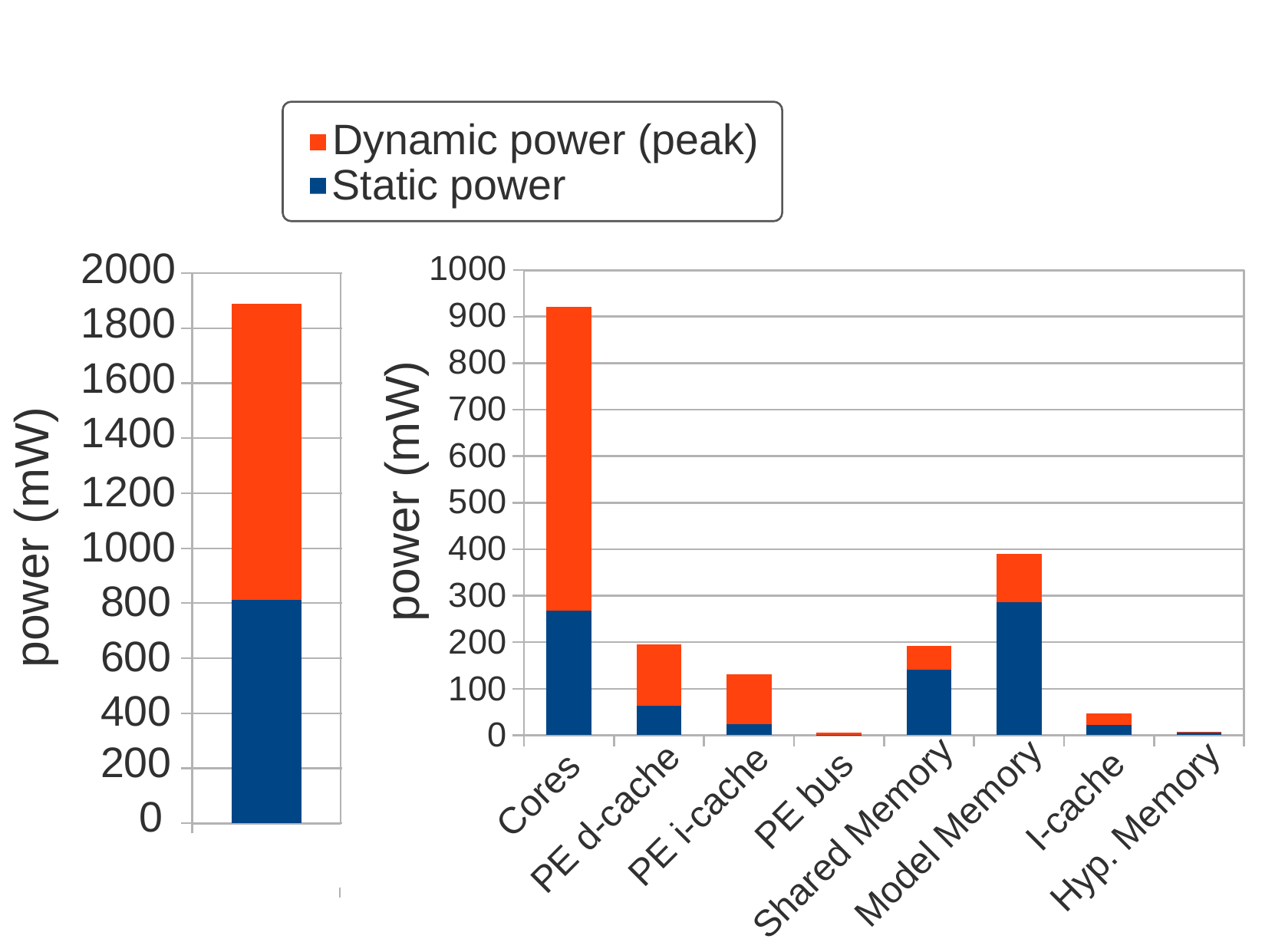}%
  \label{fig:accel_power_dyn_sta}}
    \caption{The left bar plots show the component-level breakdown of area and peak power of ASRPU. The right plots show the distribution of static and dynamic power}
  \label{fig:accel_area_ppower}
\end{figure}

Figure~\ref{fig:accel_area_ppower} shows an estimation of the area and peak power of ASRPU, broken down by component. At a 32 nm technology node, the total area is $11.68 mm^2$, $65 \%$ of which is dedicated to the execution unit (PEs, PE d-cache, PE i-cache and PE bus), 32\% is dedicated to the shared and model memories. The hypothesis unit accounts for less than $1$ \%. Regarding power, the accelerator consumes slightly more than $1.8$ W assuming peak power. That is, if every PE is in execution and every memory is accessed. Around $800$ mW come from static power, mostly from the PE cores and the shared and model memories, whereas the rest comes from dynamic power, mainly from the PE cores.

\subsection{Performance}
\begin{figure}[t]
    \centering
    \includegraphics[width=1\linewidth]{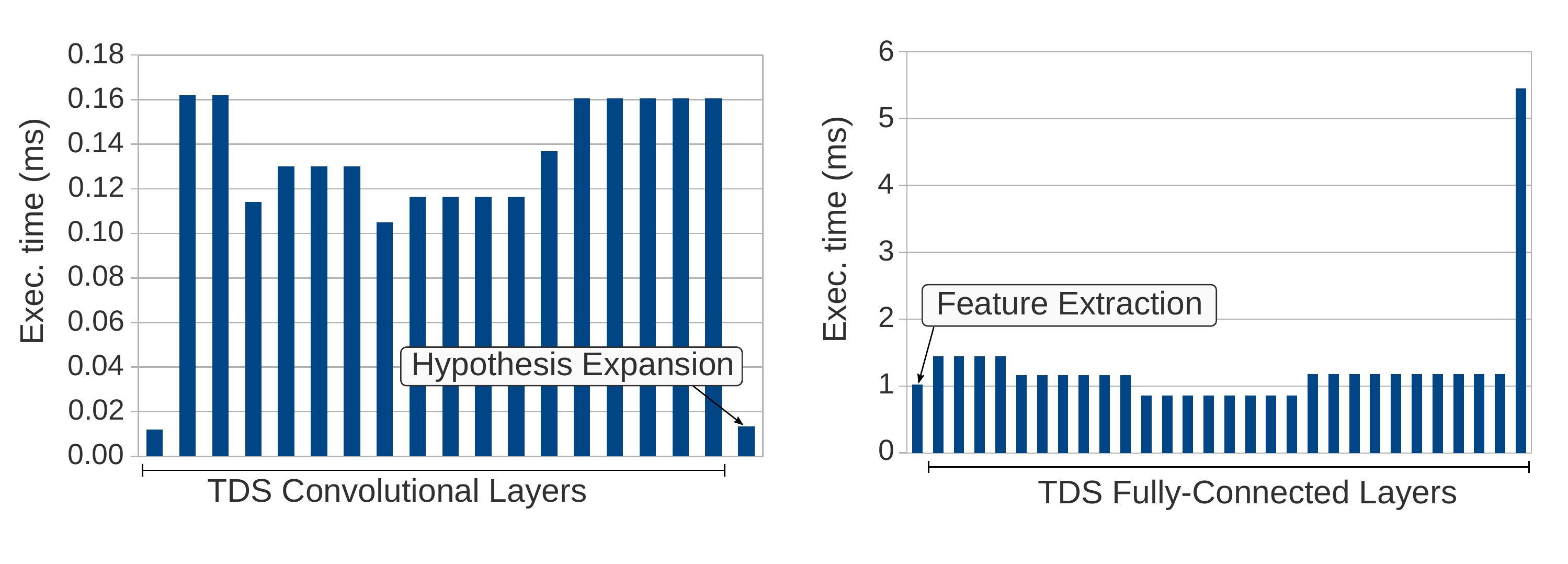}
    \caption{Execution time for the TDS ASR system running in ASRPU}
    \label{fig:tds_performance}
\end{figure}

Each decoding step in our implementation decodes $80 ms$ of audio. According to our estimations, ASRPU takes about $40 ms$ to perform a decoding step. In other words, the accelerator executes the ASR system in $2$x real-time.
Figure~\ref{fig:tds_performance} shows the execution time of the ASR system kernels, including the feature extraction and the hypothesis expansion kernels. The left plot shows the execution time for the kernels that implement the convolutional layers and the hypothesis expansion, whereas the right plot shows the execution time taken by fully-connected layers and the feature expansion during the execution of a decoding step. These estimations assume no network contention. We also assume that the model data is pre-fetched in model memory.

%Añadir algun comentario sobre los resultados de area, power y performance.

           % Results & Conclusions: 3 pages + 0 pages
\section{Conclusions}\label{sec:conclusion}

Automatic Speech Recognition is becoming a mainstream technology. Higher recognition accuracy opens the door to more use cases, while increased public acceptance towards natural human-machine interaction increases demand for ASR products. Furthermore, data privacy and Quality of service motivate an interest in performing ASR locally, on-the-edge, instead of relying on external servers, but it has to be real-time and energy efficient to be successful. 

The architecture proposed in this paper provides flexible support to implement most of the current ASR algorithms. Furthermore, given the simplicity and the few constraints imposed by the programming model on the ASR algorithms, it is likely that future algorithms can be supported, as well. The results included in this paper show how a modern ASR system can be implemented to be executed in the proposed architecture to perform real-time stream decoding with very low-power and area requirements.

There are some challenges not tackled by this work. ASR systems based on a encoder-decoder architecture (such as LAS and RNN-T) are challenging to implement in our platform. Encoder-decoder systems work as follows: The encoder part of the DNN first processes the entire sequence of features. Then, the decoder component generates score vectors until a special output is generated. This scheme does not fit well with the programming model of the proposed architecture. Supporting these systems is an interesting future work given the increasing interest in them in the ASR space. 

%% EXCLUDED FOR REVIEW <=============================================
\ifCLASSOPTIONcompsoc
  \section*{Acknowledgments}
\else
  \section*{Acknowledgment}
\fi
This work has been supported by the CoCoUnit ERC Advanced Grant of the EU’s Horizon 2020 program (grant No 833057), the Spanish State Research Agency (MCIN/AEI) under grant PID2020-113172RB-I00, the ICREA Academia program and the Spanish MICINN Ministry under grant BES-2017-080605.

%%%%%%%%% -- BIB STYLE AND FILE -- %%%%%%%%
%\newpage

\bibliographystyle{IEEEtran}
\bibliography{IEEEabrv, bibliography}
\begin{IEEEbiography}[{\includegraphics[width=1in, height=1.25in, clip, keepaspectratio]{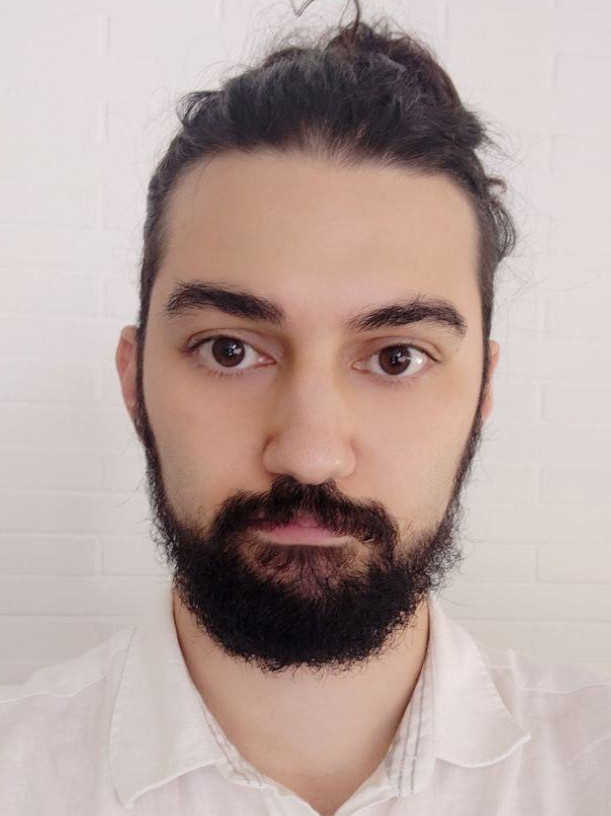}}]{Dennis Pinto}
received his BS degree in Computer Engineering in 2016 from Universidad Complutense de Madrid and his Master degree in Robotics and Automation in 2018 from Universidad Carlos III de Madrid. He is a member of the ARCO (ARchitecture and COmpilers) research group at Universitat Politècnica de Catalunya since April 2018 and is currently pursuing his PhD. His research is focused on the area of hardware support for Automatic Speech Recognition.
\end{IEEEbiography}

\begin{IEEEbiography}[{\includegraphics[width=1in, height=1.25in, clip, keepaspectratio]{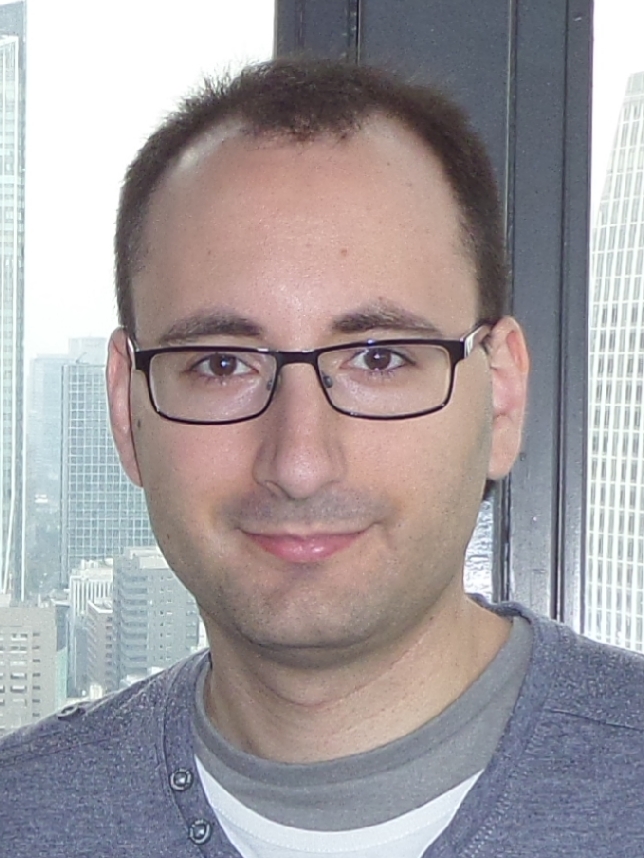}}]{Jose María Arnau}
received Ph.D. on Computer Architecture from the Universitat Politècnica de Catalunya (UPC) in 2015. He is a postdoctoral researcher at UPC BarcelonaTech and a member of the ARCO (ARchitecture and COmpilers) research group at UPC. His research interests include low-power architectures for cognitive computing, especially in the area of automatic speech recognition and object recognition.
\end{IEEEbiography}

% [width=1in,height=1.25in,clip,keepaspectratio]

\begin{IEEEbiography}[{\includegraphics[width=1in, height=1.25in, clip, keepaspectratio]{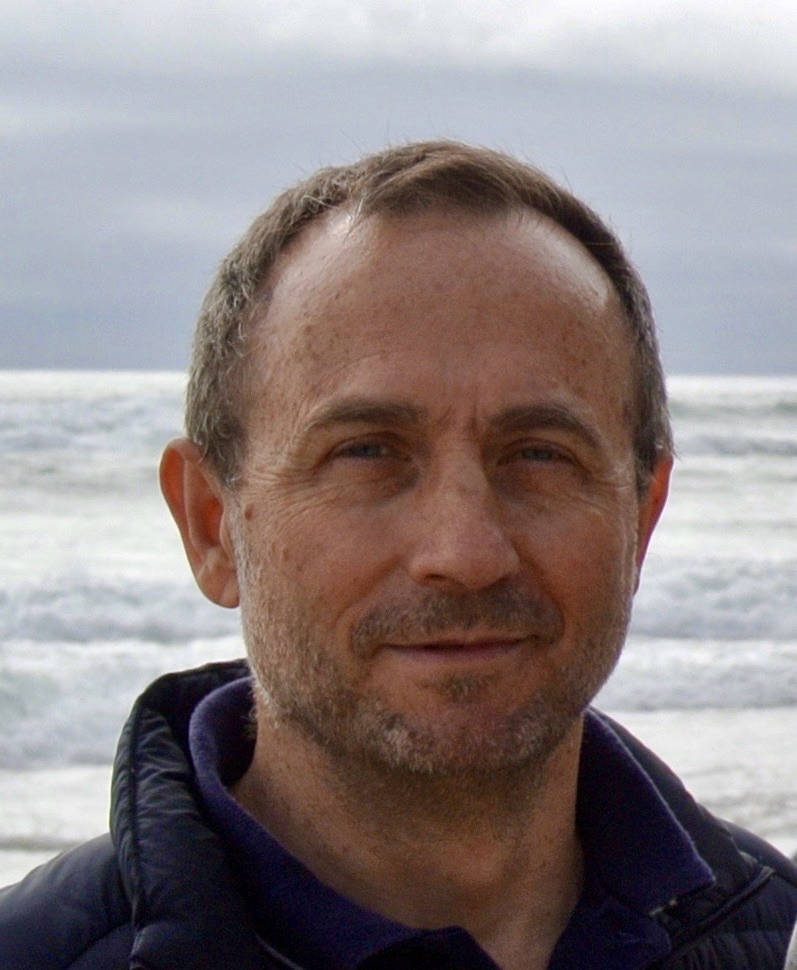}}]{Antonio Gonzalez}
(PhD 1989) is a Full Professor at the Computer Architecture Department of the Universitat Politècnica de Catalunya, Barcelona (Spain), and the director of the Architecture and Compilers research group. He was the founding director of the Intel Barcelona Research Center from 2002 to 2014. His research has focused on computer architecture and compilers, with a special emphasis on cognitive computing systems and graphics processors in recent years. He has published over 380 papers, and has served as associate editor of five IEEE and ACM journals, program chair for ISCA, MICRO, HPCA, ICS and ISPASS, and general chair for MICRO and HPCA. He is a Fellow of IEEE and ACM.
\end{IEEEbiography}

\end{document}